\documentstyle[preprint,aps,epsf]{revtex}
\begin{document}
\draft
\begin{title}
  { \bf  Shear instabilities of freely standing 
               thermotropic smectic-A films}
\end{title}

\author{Hsuan-Yi Chen and  David Jasnow}

\address{Department of Physics and Astronomy, University of Pittsburgh, 
  Pittsburgh, PA 15260, U.S.A.}

\date{\today}
\maketitle

\begin{abstract}
In this Letter we discuss theoretically the instabilities of thermotropic
freely standing smectic-A films under shear flow\cite{re:wu}.  
We show that, in Couette geometry, the centrifugal force pushes the 
liquid crystal toward the outer boundary and induces smectic layer 
dilation close to the outer boundary.  Under strong shear, this effect 
induces a layer buckling instability.  The critical shear rate is
proportional to $1/\sqrt{d}$, where $d$ is the thickness of the film.
\end{abstract}

\narrowtext

Consider shear flow experiments on freely standing fluid films in a 
Couette geometry (Fig.~1), in which the inner cylinder is 
rotating with speed $v_0$, and the outer cylinder is at rest.  
Both cylinders are perpendicular to the plane 
of the film.  When a soap film is used in the experiment, the laminar
flow between the cylinders is stable up to very large $v_0$, 
and the film gets thicker towards the outer cylinder due
to centrifugal force \cite{soap-film}.  Brightly colored patterns are 
exhibited because of this thickness variation.  However, for 
smectic-A films with smectic layers parallel to the air-fluid interface, 
which were treated as two-dimensional systems in many analyses 
\cite{re:morris}, there exists a critical speed $v_c$, such that when $v_0$ 
is above $v_c$, defects are generated \cite{re:wu}.  This clearly indicates
that the layer structure becomes important, and the system can no longer
be treated as two-dimensional.  On the other hand, 
experimental\cite{re:roux,re:safinya,re:safinya2,re:safinya3} 
and theoretical\cite{re:milner,re:bruinsma-safinya} studies on 
the rheological properties of bulk smectic-A systems show that 
under shear flow, bulk systems are most stable when the layers are
parallel to the shear plane.  Thus the finite thickness and/or the free 
surfaces have to be essential to the observed defect generation in 
the experiments on freely standing smectic-A films.

In this Letter we study the rheology of freely standing {\rm thermotropic}
smectic-A films under shear flow.  In a Couette geometry the centrifugal 
force pushes the liquid crystal away from the inner wall, which induces 
a layer dilation close to the outer wall.  
When this dilation exceeds a 
certain limit, the smectic layers are no longer stable against small 
undulational perturbations, hence defects are generated.  This instability 
is similar to the classical Helfrich-Surault type instabilities
\cite{re:clark}, but the film has free surfaces and is
far from equilibrium.  For typical materials and experimental setups
the strain induced by centrifugal force provides the critical velocity, 
$v_c$, comparable to the experimentally observed result \cite{re:wu}.  
Thus we conclude that this scenario is a good candidate for the 
mechanism which is responsible for the defect generation in 
Ref.\cite{re:wu}. Our analysis indicates that this instability occurs 
in freely standing smectic-A films but not in bulk smectic-A, since
it is a result of the interplay between crossover and boundary 
effects (2D to 3D and the existence of free surfaces), and the external 
driving flow.

\paragraph*{Strain induced buckling instability for freely standing
         smectic-A films.---}
\label{sec:strain}         
We begin with the classical strain induced buckling instability
for smectic-A in a new geometry, i.e., freely-standing films. Let the
smectic layers be parallel to the surfaces of the film, which are located
at $z=\pm d/2$, and ignore the boundaries in the $x$ and $y$ directions.
The total elastic free energy of the film is 
\cite{re:degennes-prost,re:chaikin-lubensky}
\begin{eqnarray}
F &=& 
  \alpha \left( \int dA_+ + \int dA_- \right) \nonumber \\
&&   + \frac{1}{2} \int d^2 r_{\perp} \int _{-d/2} ^{d/2} dz 
     \left(  B (E[u])^2 + K_1 (\partial _{\perp}^2 u)^2   \right) \ ,
\label{eq:nl_F}     
\end{eqnarray}     
where 
$\int dA_{\pm}=\int d^2 r_{\perp}\sqrt{1+
\left(\vec{\partial} _{\perp} [u]_{z=\pm d/2} \right)^2}$ 
is the area of the surfaces,
$u$ is the smectic layer displacement, 
$\alpha$ is the surface tension, $K_1$ and $B$ are layer
undulation and compression elastic moduli, and
$E[u]=\partial _z u -\frac{1}{2} (\vec{\partial} _{\perp} u )^2 \ $,
is the layer dilation.
The continuity of the normal stress at the surfaces 
requires~\cite{re:hy-dj-1,re:hy-dj-2,bc}
\begin{eqnarray}
 \pm \left[ B E[u] \right]_{z=\pm d/2}
= \alpha 
  (\delta A_{\pm}/\delta u)
 + P_{ext}^{\pm},
\label{eq:static_bc} 
\end{eqnarray}
where $\delta /\delta u$ stands for variational differentiation, 
$P_{ext}^{\pm}$ are the external normal stresses acting on the surface 
at $z=\pm d/2$.

Suppose the smectic layers are subject to a uniform but opposite
external stress $P_{ext}^{\pm} \equiv \pm P_{ext}$ on the surfaces; 
for small $P_{ext}$ the
layers are uniformly dilated with 
$\partial _z u= P_{ext}/B\equiv \epsilon \ $.
We study the stability of small layer undulations of
the form 
$u=u_0+u_1$
with $u_0=\epsilon z$ and 
$u_1=U N(q_y) \sin (q_yy) \cos (q_z z)$, where, for given $q_y >0$,
$q_z>0$ is chosen such that 
Eq.~(\ref{eq:static_bc}) 
is satisfied, 
and the normalization factor $N(q_y)$ is chosen such that
$\left[N(q_y)\right]^2 \int _{-d/2}^{d/2} dz \cos ^2(q_z z)=1 \ $.

Expanding $F$ to $O(U^2)$ and using the boundary condition for $u_1$ 
to the corresponding order, one finds 
\begin{eqnarray}
 F\propto \frac{1}{4} \left[B(q_z^2 - \epsilon q_y^2) 
  + K_1q_y^4\right] U^2,
\label{eq:strained-F}  
\end{eqnarray}
where, from Eq.~(\ref{eq:static_bc}), 
$q_z$ satisfies the following equation,
\begin{eqnarray}
    q_z d  = \left(l_d q_y\right)^2 \cot \frac{q_z d}{2} \ ,
\label{eq:qz}    
\end{eqnarray}
in which $l_d =\sqrt{\alpha d/B}$ is a characteristic length.
As discussed in Ref.~\cite{re:hy-dj-2},
this length characterizes the
crossover from behavior dominated by surface tension to behavior
dominated by layer elasticity in freely standing smectic-A films.    
 Since we are interested in the onset of instability,
we only consider the smallest $q_z$ which satisfies Eq.~(\ref{eq:qz})
throughout this Letter.
From Eq.(\ref{eq:strained-F}) it is clear that when the induced 
strain $\epsilon$
is large enough, there is an instability, and one would expect growth of
the amplitude $U$.

From Eq.(\ref{eq:strained-F}), the critical strain for given $q_y$ is 
$\epsilon_c (q_y)=q_z^2/q_y^2 + \lambda ^2 q_y^2 \ $, 
where $\lambda=\sqrt{K_1/B}$ is a characteristic length for 
smectic-A~\cite{re:degennes-prost}.
Defining dimensionless parameters
$  X=q_z d \ , \ Q=q_y l_d \ $,
then 
\begin{eqnarray}
  \epsilon _c (Q)
  = g \frac{\lambda}{d} \left ( \frac{X^2}{Q^2} + \frac{Q^2}{g^2} \right) 
 \equiv g \frac{\lambda}{d} T(Q,g)   \ ,
\label{eq:critical_strain}
\end{eqnarray}  
where $g=\alpha/\sqrt{K_1B}$ is a dimensionless parameter; typically 
$g\sim O(1)$.
The exact value of $X$ cannot be found analytically, but numerically
one finds that $T(Q,g)$ has a minimum at $Q=Q^*\geq 0$.  From the
definition of $T(Q,g)$, $Q^*$ depends only on $g$.
For given material parameters, one can find the minimum critical strain ,
$\epsilon ^*=\epsilon _c(Q^*)$, numerically, 
which depends only on $g$ and $\lambda/d$.  
Fig.~2
shows $Q^*$ and 
$6 \epsilon ^* d/\lambda g$ as functions of $g$. 
Notice that there is a $g_m$ such that when $g < g_m$, the 
first unstable mode has infinite wavelength.  We can estimate this
parameter by expanding the equation for $X$ around $Q=0$, and
then substituting this approximate value of $X$ into 
Eq.~(\ref{eq:critical_strain}).  We find 
$X^2/Q^2 + Q^2/g^2 = 2 + (1/g^{2}-1/3)Q^2 +O(Q^4)$.  For the first 
unstable mode to occur at $Q=0$, the coefficient of the $Q^2$ term in the
small $Q$ expansion of the critical strain has to be positive,
yielding $g_m=\sqrt{3}$. On the other hand, when 
$1/g = \sqrt{K_1B}/{\alpha} \rightarrow 0$, the surface tension is sufficiently
large such that the instability essentially becomes the Helfrich-Surault 
instability originally derived for hard boundaries, 
where the first unstable mode is
$q_{y}^*=\sqrt{\pi/d\lambda}$ and the minimum
critical strain is $2\pi \lambda/d$.~\cite{re:clark}

In the shear flow experiments conducted in Ref.~\cite{re:wu}, the freely 
standing smectic-A films were made of 
${\rm 4-cyano-4^{\prime}-octylbiphenyl}$ (8CB), which has 
$\alpha \approx 25 \ dyn/cm$~\cite{re:pon-prl}, 
$B \approx 5 \times 10^7 \ dyn/cm^2$~\cite{re:benzekri}, 
$K_1 \approx 5 \times 10^{-7} dyn$~\cite{re:vilfan}, i.e., $g \approx 5$.  
From 
Fig.~2 one finds that 
(in equilibrium), under a uniform layer dilation, the minimum 
critical strain 
$\epsilon ^*\sim \lambda g/ d =\alpha / Bd$,
and the first unstable mode has wavelength on the order of $l_d$.

\paragraph*{Shear induced instability for freely-standing smectic-A films
         in the Couette geometry.---}
The experimental setup is shown in Fig.~1.
We choose cylindrical polar coordinates to describe the system in which the
$z$ axis is the axis of the cylinders.  The radii of the inner and 
outer cylinders are $R_I$, and $R_O$, respectively.  Let the film 
be suspended horizontally on the $r\phi$ plane between the inner and outer
cylinders with thickness $d$ (from $z=-d/2$ to $z=d/2$).~\cite{typical} 
In principle $d$ depends on position in the $r\phi$ plane.
The smectic layers are parallel to the $r\phi$ plane away from the 
meniscus.  The films are usually thicker near the
meniscus due to wetting~\cite{re:sonin}, and the liquid crystal near
the meniscus acts like reservoir of material in shear flow 
experiments.~\cite{re:wu} 
Furthermore, the interactions between the liquid crystal and the walls 
also cause edge dislocations~\cite{re:pieranski}.   
Since our focus is the effect of shear flow, to avoid additional complications,
we assume both walls to  be neutral to the liquid 
crystals, and neglect the effect of gravity and the surrounding air so
that in equilibrium the smectic film is perfectly flat with a uniform 
thickness $d=d_0$.  As a result, in our model 
there is no material reservoir near the
meniscus, and the total volume of the film under shear
flow is the same as in the film at rest, a simplification
which does not occur in real experiments.

The equations of motion for this system in the isothermal, incompressible 
limit are 
\cite{re:chaikin-lubensky}
\begin{eqnarray}
    \frac{\partial u}{\partial t} + {\bf v \cdot \nabla}u
&=& v_z + \zeta _p h  \ , 
\label{eq:u} \\
  \rho \left(  \frac{\partial {\bf v}}{\partial t} + {\bf v \cdot \nabla v}
  	\right)
&=& -{\bf \nabla}p + h \hat{\bf z} + \eta \nabla ^2 {\bf v}  \ , 
\label{eq:momentum}
\end{eqnarray}
where ${\bf v}$ also satisfies the incompressibility condition
$  {\bf \nabla \cdot v} =0$.
Here ${\bf v}$ is the velocity of the liquid crystal, $\zeta_p$ is 
the permeation constant, $p$ is the pressure, and 
$h=\vec{\partial} \cdot (\delta F/\delta \vec{\partial}u)$.
In general the viscosity $\eta$ should be a second rank tensor due to the
anisotropy of the smectic-A phase \cite{re:degennes-prost,re:chaikin-lubensky}.  
However, to simplify the algebra we choose $\eta$ to be a scalar.  
Since the imposed flow direction is parallel to the smectic layers, and 
the system is isotropic in the plane defined by the layers, 
this simplification should not affect the qualitative result of 
our calculation.

The stress should be continuous across the free surfaces, which 
leads to the following conditions, \cite{bc}
\begin{eqnarray}
     \left(  \frac{\partial v_{\phi}}{\partial z}
                +\frac{1}{r} \frac{\partial v_z}{\partial \phi}
         \right)_{z=\pm d/2}                 
&=&0   \ , \
\label{eq:bc1}\\
     \left(  \frac{\partial v_z}{\partial r} 
   	       + \frac{\partial v_r}{\partial z} \right) _{z=\pm d/2}    	       
&=&0   \ , 
\label{eq:bc2}\\
    \left( -p +BE[u] \mp \alpha \ \partial _{\perp}^2 u 
           + 2 \eta \frac{\partial v_z}{\partial z} 
    \right)_{z=\pm d/2}
&=&0   \ . 
\label{eq:bc3}                   			   
\end{eqnarray}
We take no-slip boundary conditions for the velocity on the inner and 
outer walls, i.e.,
${\bf v}=v_0 \ \mbox{\boldmath $\hat{\mbox{\boldmath $\phi$}}$} 
 \   {\rm at} \ r=R_I$, and 
${\bf v}={\bf 0} \  {\rm at} \ r=R_O$.
We first consider the steady state.
Subsequently we will study the linear stability of the steady state.
We assume rotational invariance around the $z$ axis.  As a result it is 
natural to set $v_r=0$ everywhere. 
Then incompressibility and no-slip boundary conditions lead to $v_z =0$.
The momentum equation, Eq.(\ref{eq:momentum}), now becomes
\begin{eqnarray}
     \frac{{v_{\phi}}^2}{r} 
 &=& \frac{1}{\rho} \frac{\partial p}{\partial r} \ , 
 \label{eq:pressure} \\
 0 &=& \nabla ^2 v_{\phi} -v_{\phi}/r^2 
    = \frac{1}{r}\frac{\partial}{\partial r}
       (r \frac{\partial v_{\phi}}{\partial r})
       +\frac{\partial ^2 v_{\phi}}{\partial z ^2}
       - \frac{v_{\phi}}{r^2} \ , 
 \label{eq:v} \\
 0 &=& -\frac{\partial p}{\partial z} + h \ .
 \label{eq:z}
\end{eqnarray} 
The solution of Eq.(\ref{eq:v}) which satisfies the boundary conditions is
$  v_{\phi} = ar + \frac{b}{r}$,
where
$  a = - \frac{R_I}{R_O^2-R_I^2} \ v_0 \ , 
  b = \frac{R_I}{R_O^2-R_I^2} R_O^2 \ v_0 \ $.

With $v_z=0$ and $v_{\phi}$ given above, the steady state
layer displacement has to satisfy (from Eqs.(\ref{eq:u}) and~(\ref{eq:z}))
$   h=B\partial _z ^2 u - K_1 \partial _{\perp}^4 u =0$, and
$  \frac{\partial p}{\partial z} = 0$, 
i.e., the pressure depends only on $r$.
 Now one can solve Eq.(\ref{eq:pressure}) by direct integration,
leading to
\begin{eqnarray}
& &    p(r)-p(R_I)  \nonumber \\
&=& \rho \left( \frac{a^2}{2} (r^2-R_I^2) + 2ab \ln \frac{r}{R_I}
              + \frac{b^2}{2} (\frac{1}{R_I^2}-\frac{1}{r^2})
         \right)  \ ,
\label{eq:induced-p}         
\end{eqnarray}             
where $p(R_I)$ will be determined later, by conservation of total volume.
Since $p(r)-p(R_I)$ is positive definite
for non-vanishing $v_{\phi}$, the pressure near the outer
wall is larger than the pressure near the inner wall.  The boundary 
condition for the normal stress, the 
condition $h=0$, and Eq.(\ref{eq:induced-p}) together with the conservation 
of total volume will completely determine the steady state configuration.

To understand the effect of this centrifugal force induced pressure,
let us first consider the long wavelength behavior of the layer displacement
to linear order in $u$.  To this order the boundary condition for the normal
stress is
$  B\left[\partial _z u\right]_{z=\pm d/2} 
= \pm \alpha \left[\partial _{\perp} ^2 u\right]_{z=\pm d/2} 
 + \left[ p  \right]_{z= \pm d/2} \ $.
Simple dimensional analysis (replacing $\partial _z$ by $1/d$, 
$\partial _{\perp}$ by $1/(R_O-R_I)$, and
typically $B/d \gg \alpha /(R_O-R_I)^2$~\cite{typical}) shows that,
for nonvanishing $p$,
the surface tension term is negligible compared to the layer compression 
for typical experiments, and the layer dilation on the
surfaces is related to the pressure by 
\begin{eqnarray}
  B\left[\partial _z u \right]_{z=\pm d/2} 
= \left[ p \right]_{z=\pm d/2}  \ . 
\end{eqnarray}
The pressure  ``pushes'' the smectic layers on the free
surfaces, but with a very large characteristic in-plane length scale
($R_O-R_I$). 
Since the walls are neutral to the liquid crystal, and $h=0$ everywhere, 
the layer displacement is approximately
$u = \frac{p(r)}{B} \ z = \epsilon z \ $.
This layer displacement causes a change of the smectic film thickness
from $d_0$ to $d_0(1+p(r)/B)$.  Since the total volume of the liquid
crystal in the steady state is the same as that of the initial state;
we determine the pressure at $r=R_I$ from the condition
$  \int_{R_I}^{R_O} dr (2\pi r d_0)
= \int_{R_I}^{R_O} dr \left( 2\pi r d_0 \left[\frac{p(r)}{B} +1\right]
                      \right)$. 
One finds
the layers are compressed for small $r$ but are dilated for large $r$
due to the effect of centrifugal force and conservation of total 
volume.  

Consider a small patch located near the outer cylinder having size
$l \ll R_O-R_I$.  The stability to a layer undulation with wave vector in 
the radial direction can be analyzed as follows.  
Choosing $\mbox{\boldmath $\hat x$}=-\mbox{\boldmath $\hat{\phi}$}$, 
$\mbox{\boldmath $\hat{y}$}= \mbox{\boldmath $\hat{r}$}$,
when $r$ is close to $R_O$ with $R_O-r \gg l$, 
we can work in Cartesian coordinates and ignore the effect of the 
walls. The small layer undulation can be expressed as 
$U N(q_y) \sin (q_yy) \cos (q_z z)$. 
Since the shear flow is in the $x$ direction, 
the small layer undulation with in-plane
wavevector in the $y$ is not convected by the flow (i.e.,
in Eq.(\ref{eq:u}), $\vec{v} \cdot \vec{\nabla} u$ vanishes),   
and it evolves
towards $h=\vec{\nabla}\cdot \delta F/\delta \vec{\nabla} u=0$.  
Thus the equilibrium analysis can be used to
show that for large enough $v_0$, the induced layer dilation close to
the outer cylinder can 
cause a layer undulation instability. The fact that the first unstable
mode for 8CB has a wavelength on the order of $l_d$ justifies our use of
$R_O-R_I \gg l$\cite{typical}.

We are now in a position to estimate $v_c$ for the experimental geometry 
used Ref~\cite{re:wu}, i.e., $R_I=1.2$ cm, $R_O=1.5$ cm. 
Choosing $r\sim R_O$, and estimating 
$\epsilon ^*\sim \alpha / Bd$ to estimate the 
layer dilation close to the outer cylinder and the critical strain, 
one finds, for a typical material, $\rho \approx 1 \ {\rm g/cm^3}$,
$\alpha \approx 25 \ {\rm erg/cm^2}$,
$  v_c^2 \ d\sim 1500 \ {\rm cm}^3/{\rm sec}^2$. 
For $d = 3.5 \times 10^{-4}$cm, one finds
$v_c\approx 2.1 \times 10^3 {\rm cm}/{\rm sec}$, and the critical shear 
rate is $\dot{\gamma}_c = v_c/(R_O-R_I) \sim 7 \times 10^3 {\rm sec}^{-1}$.
The experimental critical shear rate is about $10^3 \ {\rm sec}^{-1}$.
Considering the simplifications made in this model, the agreement
indeed indicates that the effect of layer dilation induced by centrifugal 
force can cause the instability observed in the experiments.
Notice that our model shows the initial instability should occur close to the
outer boundary of the Couette cell, and this is consistent with the 
experimental observation.~\cite{private-wu}

\paragraph*{Concluding Remarks.---}
We have shown that freely standing smetic-A films are in principle
unstable against strong shear flow. The characteristic length $l_d$ 
introduced in Ref.~\cite{re:hy-dj-2} plays an important role. 
We considered specifically the Couette geometry and included the 
effects of centrifugal force.  We showed that centrifugal 
force is capable of inducing a layer dilation close to the outer 
cylinder.  When the
shear rate is large enough, defects are generated due to a layer
buckling instability similar to Helfrich-Surault type instability
\cite{re:clark}, and the calculated critical shear 
rate is on the same order as the experimental measurements. 

However, notice that the linear stability analysis in this Letter
is appropriate only for layer undulations with in-plane wavelength 
small compared to $R_O-R_I$.  As we point out, when 
$g=\alpha/\sqrt{K_1B} < \sqrt{3}$ the first unstable mode against 
a uniform layer dilation has very long wavelength.
Hence, our analysis does not apply to certain materials with 
weak surface tension.  For the material (8CB) used in Ref.~\cite{re:wu}, 
$g\approx 5$, and our analysis is appropriate.

Our calculations have assumed perfectly aligned smectic layers and ignored
any interactions between the air or solid boundaries with the liquid
crystal.  Edge dislocations and the material reservoir commonly observed 
near the meniscus \cite{re:pieranski}, 
which may not be negligible in the experiment, are also ignored 
in our approach.  Due to these simplifications, our work should 
only be compared semi-quantitatively with the experiments.  
We also note that, under strong shear flow~\cite{re:kellay},
defects are generated in freely standing {\rm lyotropic} 
smectic-A films in Couette geometry.  Since the density of the 
solvent is an extra hydrodynamic variable in lyotropic 
systems,~\cite{re:nallet,re:ramaswamy} 
whether the mechanism for defect generation is the same as 
that for the thermotropic systems remains to be answered.
However, our study 
has provided the basic ingredients of the experimental instability, i.e., 
the finite film thickness, surface tension, and the geometry of the 
experimental setup.  It also provides the physical picture of the 
instability, i.e., the differences 
between  a smectic-A film and a soap film in Couette geometry, and 
the differences between a bulk system and a freely standing film. 
Hence we believe that we have achieved our goal, which was to
elucidate the basic mechanism behind the interesting experiments reported
in Ref.\cite{re:wu}.  
We also add to the growing catalog of fascinating properties of 
smectic-A films far from equilibrium.

We thank D. Dash and Professor X-l. Wu for very helpful discussions.  
Professor T.~Ohta was very helpful in the early stages of this
work, and we thank him.  D.J. and H.-Y. C. are grateful for the support of
the NSF under Constract No. DMR9217935.

\newpage

\section*{Figure captions}

\noindent
Figure 1. Schematic of the experimental setup in Couette geometry,
       the $z$ axis points out of the paper
\label{fig:geometry}   
    
\noindent
Figure 2. The first unstable mode $Q^*$ and the critical strain
         $6 \epsilon ^* d/\lambda g$ 
         for freely standing smectic-A films under
          a uniform strain.
\label{fig:figure2}


\begin{references}
\bibitem{bc} The effect of permeation near the surfaces is neglected here.
             For a discussion of the permeation near the surfaces, see
             Ref.~\cite{re:hy-dj-2}.
\end{references}
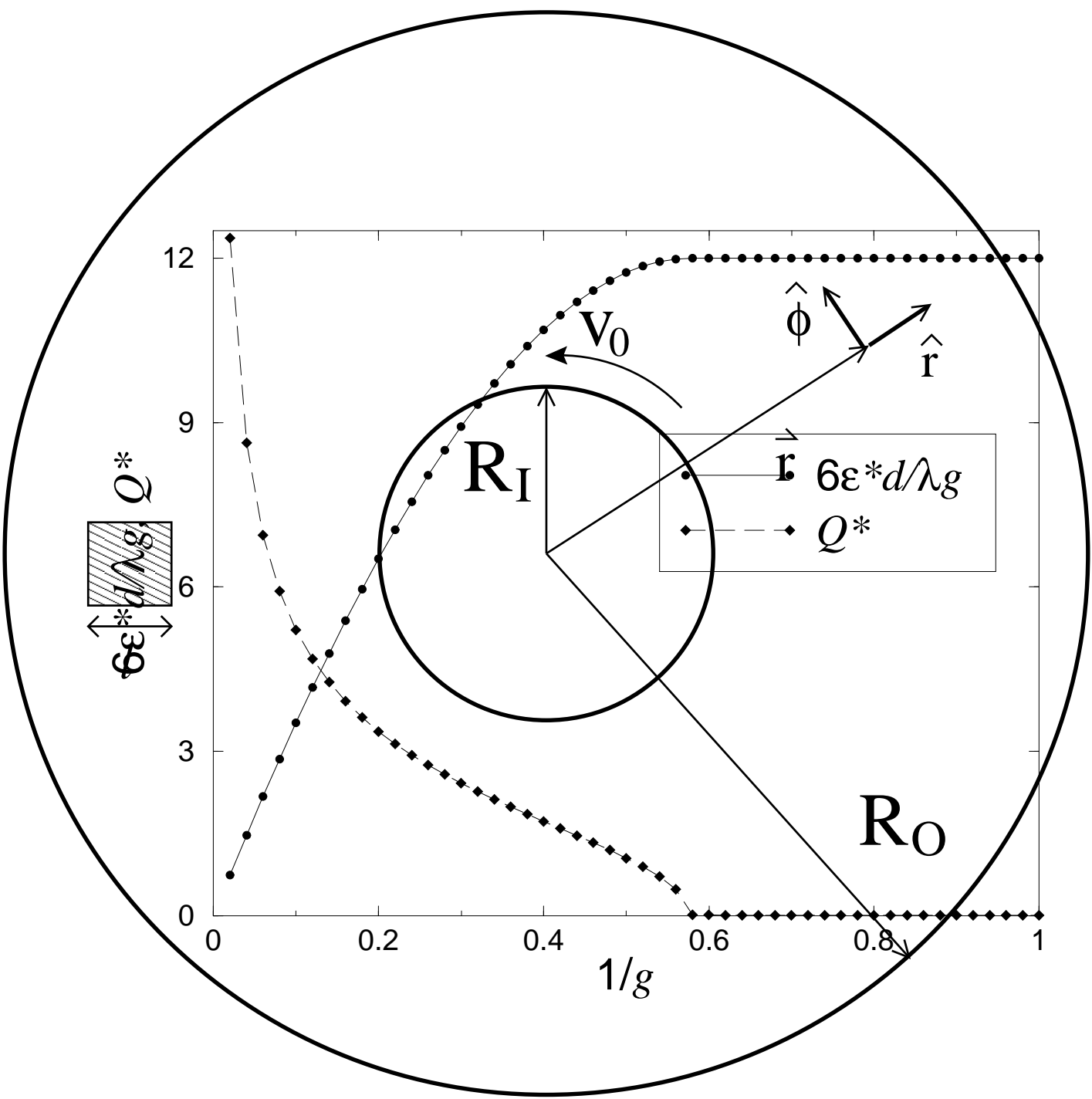
\end{document}